\documentclass{article}
\usepackage{graphics}
\begin{document}
\date{}
\title{Gravitational Lensing of Relativistic Fireball}
\author{{Eugeny Babichev and Vyacheslav Dokuchaev\thanks{
         E-mail address: dokuchaev@inr.npd.ac.ru}}\\[5mm]
         {\small\it Institute for Nuclear Research of the Russian
         Academy of Sciences}\\
         {\small\it 60th October Anniversary Prospect 7a}\\
         {\small\it 117312, Moscow, Russian Federation}}
\begin{titlepage}
\maketitle
\begin{abstract}
\baselineskip 18pt
The gravitational lensing of a relativistic fireball can produce
the time delayed multiple images with quite different spectra and
temporal patterns in contrast with a nonrelativistic source and
hence can imitate the source recurrence. In particular, the
enigmatic four multiple gamma-ray bursts detected during two days
at October 1996 in the same sky region may be due to a single
fireball event multiply imaged by the foreground galactic nucleus
or cluster of galaxies.
\end{abstract}

\vspace{8pt}
{\small {PACS numbers:} 95.30.Sf; 95.85.P; 98.62.Sb

}
\end{titlepage}

\noindent
The identical spectra and time histories of multiple images are
generally considered as necessary consequences of the gravitational
lens (GL) phenomenon (for review see {\em e.g.}~\cite{BN92}).
However these signatures of gravitational lensing are suitable only
in the case of both nonrelativistic source and GL when multiple
images are generated by the same region on the surface of the
source albeit shifted in time. Meanwhile we demonstrate in the
following that boosting of light in the relativistic source such as
a gamma-ray burst (GRB) fireball after an appropriate lensing may
provide the multiple images without any similarities of spectra and
light-curves.

The observed cosmological GRBs are the promising targets for the
gravitational lensing of relativistic sources. The similarities of
spectra and temporal patterns of multiple images were used until
now as observation signatures in searches of the possible
gravitational lensing of GRBs with a hope to determine or confine
{\em e.g.\/} the contents and density of compact objects in dark
matter and average redshift of GRBs \cite{BW92,HMQ99,Marani99}. The
GRB, within the framework of a standard cosmological scheme of
their origin, is generated by the nearly spherical or beamed low
mass leptonic relativistic fireball resulting from the coalescence
of tight neutron star binary or collapse of some massive star in
distant galaxy (for reviews see {\em
e.g.\/}~\cite{Piran99,Postnov99}). A typical expected recurrence
rate of cosmological GRBs is one per million years per galaxy.
Hence if possible observation of GRB repetition were not due to the
time delayed multiple lensed images, it would create hard problems
for the standard cosmological scheme of GRB origin. The hunting for
GRB recurrence is the difficult task because of the low angular
resolution of nowadays gamma-telescopes. Recent statistical
analysis of GRB samples provide only the upper limits on the
possible repetition of GRBs \cite{BR96,Lamb96,Tegmark96}. Meanwhile
among more than 2500 GRBs detected nowadays there are surprising
multiple GRBs coming at October 1996. At that time the orbital
telescopes BATSE, KONUS, TGRS and Ulysses detected independently
four GRB events from the same sky region during two days
\cite{Meegan96,Conn97}. The probability of accidental projection of
these four GRBs is very small,
$3.1\cdot10^{-5}\div3.3\cdot10^{-4}$, whereas the clustering is not
so significant if the four events combine into the three bursts
\cite{GrazLQ98}. However in the latter case one of the bursts would
be more than 20 minutes in duration. Nevertheless the recurrence of
cosmological GRBs is a rather natural if some part of GRBs are
generated not by the coalescence of neutron star binaries but by
the accidental collisions of neutron stars in the dense central
stellar clusters of distant galactic nuclei \cite{DEO98}. In this
case the observation of multiple GRBs is a consequence of a
serendipitious event of massive black hole formation accompanied by
the four events of neutron star collisions during dynamical
collapse of extremely dense central stellar cluster in some distant
galactic nucleus \cite{DEO97}.

\begin{figure}
\resizebox{\hsize}{!}{\includegraphics{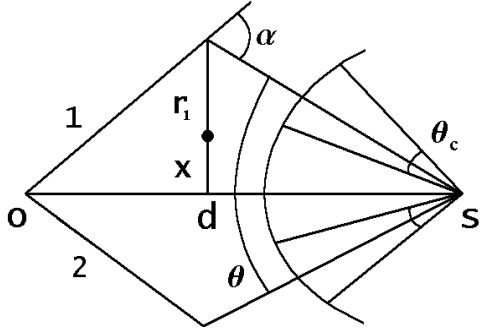}}
\caption{Schematic view of the gravitational lensing of
relativistically expanding fireball with a center of deflector (d)
shifted to distance $x$ from the line connecting the source (s) and
observer (o) in the absence of deflector. Rays 1 and 2 leave the
fireball with a separation angle $\theta$ and have correspondingly
the impact parameters $r_1$ and $r_2$ which define deflection angle
$\alpha$. Relativistic boosting confines each visible image by the
narrow region of the expanding fireball within an open angle
$\theta_c=\Gamma^{-1}\ll1$ relative to the fireball center.}
\label{fig1}
\end{figure}

Here we propose the other resolution of the `fast GRB recurrence'
problem. We demonstrate in the following that a cluster of four GRB
events observed during two days at October 1996 can be all due to a
single relativistic fireball event at cosmological distance
expanding with a large bulk Lorentz factor $\Gamma\gg1$ and
multiply imaged by the GL. The basic idea of specific gravitational
lensing with resulting different spectra and temporal patterns of
separate time delayed images is taking into account the strong
aberration (boosting) of light from the relativistically expanding
fireball. This boosting confines the viewed portion of the
expanding fireball by the narrow confinement angle
$\theta_c\sim\Gamma^{-1}\ll1$ relative to the fireball center (see
Fig.~\ref{fig1} for details).

The emission of the fireball beyond the confinement angle
$\theta_c$ comes to the detector without the boosting and generally
sinks down to the noise. In this case two `rays' separated by the
angle $\theta>\theta_c$ originate from the different parts of the
fireball connected by the space-like world lines. These regions are
causally disconnected and their emission may be generated under
different physical conditions. An additional supposed requirement
for the realization of this model is a small-scale turbulent
structure of emitting shocks in the fireball with a typical length
much less than the instant fireball radius. In this case the
outgoing rays would not retain information on the temporal
structure of the central source of energy. This highly turbulent
structure seems quite reasonable for the GRB fireball because of
developing of instabilities in the relativistic shocks (see {\em
e.g.\/} \cite{Piran99}).

Gravitational lensing of {\em nonrelativistic} source provides
multiple images only with time-shifted but similar spectral and
temporal structures. On the contrary outgoing rays from the {\em
relativistic} fireball separated by the angle $\theta>\theta_c$ in
general can be generated under the different physical conditions.
After an appropriate gravitational lensing they would appear to the
distant low-resolution telescope as recurrent separate events from
the same prospective source on the sky but with the quite different
spectral and temporal structures. This specific feature of the
gravitational lensing of relativistic fireballs may influence the
results of statistical searches of GRB repetition of
\cite{BR96,Lamb96,Tegmark96}. In the following we determine the
necessary physical parameters of the possible GL to imitate the
recurrent multiple GRBs with different spectra and light curves
from the same GRB event.

For a general mass distribution in the GL (deflector) with the
Newtonian potential $\phi(\vec{r})$ the deflection angle of a
separate ray can be expressed (see {\em e.~g.} \cite{BK96}) as a
two dimensional vector
\begin{equation}\label{vecalpha}
\vec{\alpha}=\frac{2}{c^2}
\int ds \;\vec{n} \times(\vec{n}\times{\mathbf\nabla}\phi),
\end{equation}
where $\vec{n}$ is a unit vector along the ray and the integral is
performed along the ray too. Fig.~\ref{fig1} represents the
schematic view of the possible GRB ({\it i.~e.} relativistic
fireball) lensing geometry with a center of GL (deflector) in
general shifted from the line connected the observer and source
which is the same as the path of the undeflected ray ({\em i.~e.}
in the absence of deflector). The estimation for potential is
$\phi\sim\sigma^2$ for the GL composed of constituent masses ({\em
e.~g.} stars or galaxies in the cluster) moving with a virial
velocity dispersion $\sigma$. It follows then from
Eq.~(\ref{vecalpha}) that a value of the deflection angle
$\alpha=|\vec{\alpha}|$ is always small,
$\alpha\sim\phi/c^2\sim\sigma^2/c^2\ll1$. In the limit of a small
deflection angle there are general relations for a geometric time
delay $\Delta t_g$ which is conditioned by the path difference
between the deflected and undeflected rays
\begin{equation}
\label{deltag}
\Delta t_g=(1+z_d)\frac{\alpha\xi}{2c}
\end{equation}
and for a separation $\xi$ between deflected and deflected rays
\begin{equation}
\label{xi}
\xi=\frac{D_d D_{ds}}{D_s}\alpha,
\end{equation}
where $D_s$, $D_d$, and $D_{ds}$ are the angular diameter distances
\cite{Weinberg} from the observer to the source of GRB, to the
deflector (lens) and from the deflector to the source
correspondingly and $z_d$ is a deflector redshift. Besides the
geometric time delay $\Delta t_g$ there is an additional general
relativistic time delay $\Delta t_{gr}$ due to traversing through
the region with a gravitational field (``Shapiro delay'')
\begin{equation}
\label{deltagr}
\Delta t_{gr} =-\frac{2}{c^3}\int ds\;\phi.
\end{equation}
The both geometrical $\Delta t_g$ and gravitational $\Delta t_{gr}$
time delays are of the same order of magnitude for the general ray
position as can be verified from Eqs.~(\ref{deltag}) and
(\ref{deltagr}) and using estimation $\alpha\sim\phi/c^2$ from
Eq.~(\ref{vecalpha}). So the corresponding total time delay $\Delta
t=\Delta t_g + \Delta t_{gr}\sim GM(\xi)/c^3$ is defined by the
effective mass $M\sim M(\xi)$ of the GL inside the radius $r=\xi$.
It must be taken in mind that in some degenerate cases of GL
symmetry this estimation of time delay between images may provide
only the upper limit on $\Delta t$ because the time delays of
different rays tend to compensate each other.

Now we can formulate two necessary requirements for production of
lensing images of causally disconnected regions of relativistic
fireball. For the general case of both the lens (deflector) and
relativistic fireball (source) at comparable (cosmological)
distances $D_d\sim D_s$ the angle between two outgoing rays is
$\theta=\alpha (D_d/D_s)\simeq\alpha$. Using deflection angle
estimation $\alpha\sim\sigma^2/c^2$ and causal disconnection
condition $\theta>\theta_c\sim\Gamma^{-1}$ we find the first
requirement on the velocity dispersion $\sigma$ in the GL:
$\sigma\geq c/\Gamma^{1/2}$ with expected $\Gamma\sim10^3$ for GRB
case. The second requirement on the effective total mass $M$ of the
GL follows from the estimation of the time delay $\Delta
t(1,2)=\Delta t(1)-\Delta t(2)$ between two images: $M\sim
c^3\Delta t(1,2)/G$ with $\Delta t(1,2)\sim1$~day. One of this
requirements may be replaced by the equivalent one for the GL
radius $R<\Gamma c\Delta t(1,2)$ with the use of virial relation
$R\sim GM/\sigma^2$.

To specify more explicitly these requirements we consider a simple
model of the gravitational lensing of relativistic fireball by the
singular isothermal sphere (SIS) \cite{GottG74,Young80} with a
center of SIS shifted to distance $x$ from the line connecting the
source and observer (which is the path of undeflected ray) as shown
on Fig.~\ref{fig1}. It is convenient to suppose that SIS is
confined within some finite radius $R$. The spherically symmetric
and constant temperature mass distribution $M(r)$ in this model of
GL is connected with a current radius $r$ by the relation
\begin{equation}\label{radius}
r=\frac{GM(r)}{\sigma^2},
\end{equation}
where $\sigma=const\ll c$ is a line-of-sight velocity dispersion in
the SIS. The deflection angle of this GL is independent on the
impact parameter of the ray (while it is much less than $R$) and
equals
\begin{equation}
\label{alpha}
\alpha=4\pi{\left(\frac{\sigma}{c}\right)}^2.
\end{equation}
This Eq. specifies the first requirement for causal disconnection
of images, $\alpha\geq\Gamma^{-1}$, which provides the restriction on
the velocity dispersion in the GL:
\begin{equation}\label{sigma}
\sigma\geq\frac{c}{(4\pi\Gamma)^{1/2}}\simeq3\cdot10^3
\Gamma_{3}^{-1/2} \mbox{~km s}^{-1},
\end{equation}
where $\Gamma_{3}=\Gamma/10^3$.

To determine the required scales of time delay $\Delta
t(1,2)\sim1$~day and required deflection angle $\alpha$ we consider
two images produced by the lensing of two rays, `ray~1' and
`ray~2', coming through opposite sides of the SIS with
corresponding impact parameters $r_1$ and $r_2$ such that
$\max(r_1,r_2)\ll R$. The gravitational potential inside the SIS
(at $r<R$) is
\begin{equation}
\label{potential}
\phi(r<R)=-\sigma^2 \left(1+\ln\frac{R}{r}\right).
\end{equation}
Because of $\alpha=const$ any two rays coming through opposite
sides of the SIS have the same separation $\xi_1=\xi_2=\xi=(r_1 +
r_2)/2$. As a consequence the geometric time delay between ray 1
and ray~2 equals zero according to Eq.~(\ref{deltag}). So the only
time delay between considered two rays results from the difference
of gravitational time delays $\Delta t(1.2)=\Delta t_{gr}(1)-\Delta
t_{gr}(2)$. After the simple but lengthy calculations we find
\begin{equation}
\label{delay}
\Delta t(1,2)=4\pi(1+z_d)\left(\frac{\sigma}{c}\right)^2\frac{x}{c}
=4\pi(1+z_d)\frac{GM(x)}{c^3},
\end{equation}
where the last equality follows from Eq.~(\ref{radius}). According
to limitation of Eq.~(\ref{sigma}) the shift of SIS center
$x=(r_2-r_1)/2$ is
\begin{equation}
\label{x}
x\leq\frac{c\Delta t(1,2)}{1+z_d}\Gamma\simeq
\frac{0.84}{1+z_d}\;\frac{\Delta t(1,2)}{1\mbox{~day}}\;
\Gamma_3\mbox{~pc}.
\end{equation}
Using $\Delta t(1,2)$ from Eq.~(\ref{delay}) as an observed time
delay between different GRB lensing images of the same relativistic
fireball we find the required mass of the GL:
\begin{equation}\label{mass}
M\geq M(x)\simeq\frac{1.4\cdot10^{9}}{1+z_d}\;\frac{\Delta
t(1,2)}{1\mbox{~day}}\;{\rm M_{\odot}};
\end{equation}
This estimation with an order of magnitude accuracy is valid for
the more complex mass distributions inside the GL and in fact
provides only the low limit of the possible GL mass because of the
possible time delay compensation between different images.

It follows from Eqs.~(\ref{sigma}) and (\ref{mass}) that the
necessary restrictions on the parameters of GL which produce (i)
the one day time delayed and (ii) causally independent images (with
different spectra and temporal histories) of the same single GRB
fireball are the following: the required velocity dispersion of the
constituent masses in the GL for the generation of multiple GRB
events at October 1996 is $\sigma\geq3\cdot10^3$~km~s$^{-1}$ and
the total mass of the GL $M\geq10^9{\rm M_{\odot}}$ for the
effective redshift of GRBs $z\simeq1\div2$. These required values
of $M$ and $\sigma$ may be realized in the real case of lenses such
as a dense enough cluster of galaxies or massive central stellar
cluster in galactic nucleus. In addition Eqs.~(\ref{xi}) and
(\ref{sigma}) provide another restriction: for a rather general
case of $D_s\sim D_d$ the deflector must be situated at the
distance from the source
$D_{ds}\sim\xi/\alpha\leq10^{3}\Gamma_{3}R$. For the angular
diameter distance $D_{ds}$ of cosmological scale the suitable
$R\sim1$~Mpc is a characteristic scale for the cluster of galaxies
and respectively for $D_{ds}\sim1$~kpc (GRB originated in distant
galactic disk) the suitable $R\sim1$~pc is a scale of the central
stellar clusters in the host galactic nucleus. These both examples
of promising GL candidates are again in agreement with other
previous requirements. At the same time these both deflectors are
the most common examples of the nowadays observed cosmological
lenses. Meanwhile the considered model works for fireballs emerging
inside (or closely) to deflectors: whether (i) in one of the
galaxies in the host cluster or (ii) in the galactic disk near the
host galactic nucleus. This proximity of the source to the
deflector increases the probability of the lensing with respect,
for example, to the lensing of quasars (by distant foreground
galaxy or cluster of galaxies).

The gravitational potential of real GL must be nonspherical to
generate 4 or more images. The simplest model is the SIS potential
perturbed by a quadratic share \cite{BK88} {\em e.g.\/} caused by
rotation, which in general generates one faint and four bright
images. Massive rotating black hole can also produces the suitable
nonspherical gravitational field.

In general we demonstrate that identities of spectra and time
histories cannot serve as specific signatures of gravitational
lensing in the case of relativistic sources. In particular the
gravitational lensing of relativistic fireball from a single GRB
event may imitate the GRB recurrence.

\end{document}